\documentclass[twocolumn,amsmath,amssymb]{revtex4}

\usepackage{graphicx}
\usepackage[latin1]{inputenc}
\usepackage{psfrag}

\newcommand{\eq}[1]{\begin{equation} #1 \end{equation}}
\newcommand{\eqa}[1]{\begin{eqnarray} #1 \end{eqnarray}}
\newcommand{\sss}{\scriptscriptstyle}
\def\lg{^{\rm long}}
\def\Amix{\mathcal{A}_{\rm mix}}
\def\Adir{\mathcal{A}_{\rm dir}}

\begin{document}

\title{\bf  Evading $1/m_b$-suppressed IR divergencies in QCDF:\\
$B_s\to KK$ Decays and $B_{d,s}$ mixing\footnote{Talk given at the International
Workshop on Quantum Chromodynamics: QCD@Work 2007, Martina Franca, Italy, June 2007.}}

\author{Javier Virto\\\vspace{0.6cm}}
\affiliation{Grup de F{\'\i}sica Te{\`o}rica and IFAE,\\ Universitat Autònoma de \mbox{Barcelona, 08193} Bellaterra, Spain}

\begin{abstract}
We analyze the deviations of the mixing induced CP asymmetry in $B\to \phi K_s$ from $\sin{2\beta}$, as well as the deviations
of the asymmetries in $B_s\to K^{*0}\bar{K}^{*0}$, $B_s\to \phi K^{*0}$ and $B_s\to \phi\phi$ from $\sin{2\beta_s}$, that arise
in SM due to penguin pollution. We use a theoretical input which is short-distance dominated in QCD-factorization and thus free of IR-divergencies.
We also provide alternative ways to extract angles of the unitarity triangle from penguin-mediated decays, and give predictions
for $B_s\to K^{*0}\bar{K}^{*0}$ observables.
\end{abstract}

\maketitle

\section{Introduction}

The phenomenology of hadronic $B_d$-decays has been a matter of intensive research in the past 15 years, in part due to the large amount of data
collected at the B-factories Babar and Belle, and at CDF and D0. This research has led to the consensus that the CKM mechanism
for CP and flavor violation is accurate. However, some puzzles have survived up to this day \cite{BpiKex,BpiKth,BVKex,BVKth,Stewart}.

A new era in B-physics has been triggered by the experimental precision and by the emergent exploration of the $B_s$ system. The future is
marked by the starting of LHCb and the possibility of a super-B factory. So far, measurements on the $B_s$ system are limited
to the mass difference $\Delta M_s$ \cite{DMs}, other mixing parameters \cite{mix}, and some $B_s\to \pi K$ and $B_s\to KK$ modes
\cite{Bsmodes}, but the extension of this list is an important element in the B-physics program.

On the theoretical side, the study of non-leptonic $B$-decays is difficult because of the presence of important long distance strong interaction
effects. The correct computation of these contributions is crucial to be able to resolve small NP contributions.
The amplitude of a B meson decaying into two light mesons can be written as
\eq{A(B\to M_1M_2)=\lambda_u^{(D)*}\,T_{M_1M_2}+\lambda_c^{(D)*}\,P_{M_1M_2}}
were $\lambda_q^{(D)}\equiv V_{qb}V_{qD}^*$, and $T$ and $P$ are called ``tree'' and ``penguin''. These hadronic parameters can be extracted
from data, or can be predicted using symmetries (such as flavor). The direct computations from QCD are much more involved and are
based on factorization and the $1/m_b$ expansion. They appear in the context of QCDF, pQCD or SCET \cite{fact}. While the methods based
on flavor symmetries include naturally all kinds of long distance physics, they suffer from big uncertainties due to bad data
an poorly estimated SU(3) breaking. On the other hand, methods based on factorization suffer from uncertainties due to non-factorizable chirally
enhanced $1/m_b$ corrections, and long distance charm loops (charming penguins).

These proceedings review a recent proposal to improve (at a phenomenological level) on some of the weak points
of the approaches mentioned above \cite{DMV,BVV}. We also include a straightforward application to $B\to \phi K_s$ and
comment on the limitations and the applicability of the method.

\section{Express review of $B_q-\bar{B}_q$ mixing}

%
%
%
%
%
%

The time evolution of a $B_q$ meson ($q=d,s$) can be easily described by changing to the mass eigenbasis. The relationship
between the flavor basis and the physical mesons is specified by a mixing parameter usually denoted by $q/p$,
\eqa{
|B_L\rangle&=&\frac{1}{\sqrt{1+|q/p|^2}}\Big( |B^0\rangle + \frac{q}{p}\,|\bar{B}^0\rangle\Big)\nonumber\\
|B_H\rangle&=&\frac{1}{\sqrt{1+|q/p|^2}}\Big( |B^0\rangle - \frac{q}{p}\,|\bar{B}^0\rangle\Big).
}
The time evolution of the mass eigenstates is straightforward and depends only on the masses and the widths
of the physical B-mesons. Therefore, the evolution of the flavor eigenstates --which describes the oscillations--
is specified by the masses and widths of the physical mesons and on the mixing parameter $q/p$.

Then one can define the mixing angle $\phi_M$ as
\eq{\phi_M\equiv -\arg{(q/p)}}
In terms of the entries of the effective hamiltonian,
\eq{\frac{q}{p}=\sqrt{\frac{M_{12}^*-\frac{i}{2}\Gamma_{12}^*}{{M_{12}-\frac{i}{2}\Gamma_{12}}}}\simeq \sqrt{\frac{M_{12}^*}{M_{12}}}}
where we have used that $|\Gamma_{12}|\ll |M_{12}|$. This means that the above definition of the mixing angle is equivalent to
\eq{\phi_M=\arg{(M_{12})}}
One should be aware, though, that these quantities are not convention independent and are sensitive to unphysical phase redefinitions.
However, once a convention for the weak phases is chosen everywhere, this definition is meaningful. In the Wolfenstein parametrization
$V_{ub}$ and $V_{td}$ have phases of $\mathcal{O}(1)$ and $V_{ts}$ of $\mathcal{O}(\lambda^2)$. This means that in SM, in the case of $B_d-\bar{B}_d$
mixing $\ M_{12}^d\propto (V_{td}^*V_{tb})^2$ and $\phi_d^{SM}=2\beta+\mathcal{O}(\lambda^4)$, and in the case of $B_s-\bar{B}_s$
mixing $\ M_{12}^s\propto (V_{ts}^*V_{tb})^2$ and $\phi_s^{SM}=2\beta_s$.

In order to measure the mixing angle $\phi_M$ one usually looks at CP asymmetries. The time-dependent CP asymmetry of a $B(t)\to f$ decay
is defined as
\eq{\mathcal{A}_{\rm CP}(t)\equiv\frac{\Gamma(B(t)\to f)-\Gamma(\bar{B}(t)\to \bar{f})}{\Gamma(B(t)\to f)+\Gamma(\bar{B}(t)\to \bar{f})}}
where $B(t)\ (\bar{B}(t))$ was a $B\ (\bar{B})$ at $t=0$. For $f=f_{\rm CP}$ a final CP-eigenstate (with CP eigenvalue $\eta_f$),
and neglecting the small CP violation in mixing one finds
\eq{\mathcal{A}_{\rm CP}(t)=\frac{\Adir\cos(\Delta M t)+\Amix\sin(\Delta M t)}
{\cosh(\Delta \Gamma t/2)-\mathcal{A}_{\Delta\Gamma}\sinh(\Delta\Gamma t/2)},}
where the quantities $\Adir$ and $\Amix$ are the so called direct and mixing induced CP asymmetries, and are measured
from the time oscillations of $\mathcal{A}_{\rm CP}(t)$. The mixing induced CP asymmetry is given by
\eq{\Amix=-\frac{2{\rm Im}\lambda_f}{1+|\lambda_f|}\ ;\quad \lambda_f\equiv \frac{q}{p}\frac{\bar{A}_f}{A_f}\simeq e^{-i\phi_M}\frac{\bar{A}_f}{A_f}}
where $\lambda_f$ is a convention-independent (physical) quantity. Here $\bar{A}_f\equiv A(\bar{B}\to f)$
and $A_f\equiv A(B\to f)$.

In some particular cases, one can extract the mixing angle in a very clean way from
a mixing induced CP asymmetry. The prominent example is the case of $B_d\to J/\psi K_s$ \cite{jpsi0}.
Since this decay is dominated by a single amplitude, $\bar{A}_{J/\psi K_s}/A_{J/\psi K_s}\simeq \eta_{J/\psi K_s}=-1$,
and therefore,
\eq{-\Amix(B_d\to J/\psi K_s)\simeq \sin{2\beta}}
The neglected amplitude is both CKM and $\alpha_s(m_b)$ suppressed with respect to the dominant amplitude,
so the corrections to this equation are below the percent (or even the per mil) level \cite{jpsi1,jpsi2,jpsi3}.

The same argument is valid in principle for penguin-mediated $b\to s$ decays, where the amplitude
can be written as
\eq{A=\lambda_u^{(s)*}\,T+\lambda_c^{(s)*}\,P\,\stackrel{!}{\simeq}\, \lambda_c^{(s)*}\,P.}
The tree part is strongly CKM suppressed since $|\lambda_u^{(s)}/\lambda_c^{(s)}|\sim 2\%$. However,
as opposed to the previous case, $T$ is not suppressed with respect to $P$. In fact, an enhancement of
$T$ over $P$ would spoil the extraction of $\phi_M$ from these modes.

The classical example is the case of $B_d\to \phi K_s$, where
\eq{-\Amix(B_d\to \phi K_s)=\sin{2\beta}+\Delta S_{\phi K_s}}
with
\eq{\Delta S_{\phi K_s}=
2\left| \frac{\lambda_u^{(s)}}{\lambda_c^{(s)}} \right|{\rm Re}\left( \frac{T_{\phi K_s}}{P_{\phi K_s}} \right)\sin{\gamma}\cos{2\beta}+\cdots}
In order to have under control the smallness of the correction $\Delta S_{\phi K_s}$ one should be able to bound the
size of ${\rm Re}(T_{\phi K_s}/P_{\phi K_s})$. In fact, latest data gives \cite{HFAG}
\eq{\Delta S_{\phi K_s}^{\rm exp}=-0.39\pm 0.20}
so if the uncertainties are reduced around this central value, the claim of a NP signal will have to rely on a
solid bound for the SM penguin pollution.

A first approach to bound this tree-to-penguin ratio was taken in the context of flavour SU(3) symmetry \cite{phiKs1,phiKs2}. The argument is that
if there was a large hierarchy between T and P in a $b\to s$ mode, the hierarchy would persist when one moves to the SU(3)-related
$b\to d$ modes. But in these modes the tree is not suppressed with respect to the penguin, because $|\lambda_u^{(d)}|\sim |\lambda_c^{(d)}|$,
so this hierarchy would have a clear impact in the observables of the $b\to d$ modes. In this way one can write the following bound
\eq{|\Delta S_{\phi K_s}|<\sqrt{2}\lambda
\Bigg( \sqrt{\frac{BR_{\phi \pi^+}}{BR_{\phi K_s}}}+ \sqrt{\frac{BR_{K^{*0}K^+}}{BR_{\phi K_s}}}\Bigg)+\mathcal{O}(\lambda^2)}
valid in the SU(3) limit and under a non-cancelation assumption between $B_d\to \phi K_s$ and $B^+\to \phi K^+$. With the present data
the bound is
\eq{|\Delta S_{\phi K_s}^{SU(3)}|<0.4}
The same kind of analysis can be applied to other penguin-dominated decays \cite{phiKs2,othersin2beta}.

A second approach has been followed in the framework of QCD-factorization \cite{beneke}, which gives a much more
competitive bound,
\eq{0.01<\Delta S_{\phi K_s}^{\rm QCDF}<0.05}
Related analyses have been carried out in SCET \cite{zupan} and pQCD \cite{mishima}. For a recent review see also \cite{ZupanTalk}.

Concerning the $B_s-\bar{B_s}$ mixing angle, the clean tree-level determination comes from $B_s\to J/\psi \phi$.
The related penguin-mediated decays are, for example, $B_s\to K^{*0}\bar{K}^{*0}$, $B_s\to \phi K^{*0}$ or $B_s\to \phi\phi$.
Again, one can write
\eq{\eta_f\Amix(B_s\to f)=\sin{2\beta_s}+\Delta S_{f}.}
A study of the amounts by which their mixing induced CP asymmetries deviate from
$\sin{2\beta_s}$ in the SM can be found in \cite{BsSilvestrini,BVV}.

In the next pages we follow the approach in \cite{BVV,DMV}, based on a theoretical
input that we call $\Delta$.

\section{Theoretical input}

Consider the quantity $\Delta\equiv T-P$. This quantity is a hadronic, process-dependent,
intrinsically non-perturbative object, and thus difficult to compute
theoretically. Such hadronic quantities are usually either extracted
from data or computed using some factorization-based approach. In
the latter case, $\Delta$ could suffer from the usual problems
related to the factorization ansatz and in particular long-distance
effects.

However, for a certain class of decays, $T$ and $P$ share the same
long-distance dynamics: the difference comes from the ($u$ or $c$)
quark running in the loop, which is dominated by short-distance
physics~\cite{DMV}. Indeed, in such decays, $\Delta=T-P$ is not
affected by the breakdown of factorization that affects annihilation
and hard-spectator contributions, and it can be computed in a
well-controlled way leading to safer predictions and smaller
uncertainties.

\begin{table}
\begin{tabular}{lcc}
\hline\hline
$\Delta_{\phi K_s}^d$ & $\quad$ &$(2.29\pm 0.67)\times 10^{-7}{\rm GeV}$\\
\hline
$\Delta_{K^*K^*}^d$ & &$(1.85\pm 0.79)\times 10^{-7}{\rm GeV}$\\
\hline
$\Delta_{K^*K^*}^s$ & &$(1.62\pm 0.69)\times 10^{-7}{\rm GeV}$\\
\hline
$\Delta_{\phi K^*}^s$ & &$(1.16\pm 1.05)\times 10^{-7}{\rm GeV}$\\
\hline
$\Delta_{\phi\phi}^s$ & &$(2.06\pm 2.24)\times 10^{-7}{\rm GeV}$\\
\hline\hline
\end{tabular}
\caption{Values of $\Delta$ for the various decays of interest. In the case of
two vector mesons these numbers correspond to longitudinal polarizations.}
\label{table1}
\end{table}

Table \ref{table1} shows the values of $\Delta$ for our cases of interest. This quantity was used
to predict branching ratios and asymmetries in $B_s\to KK$ modes, and the outcome was promising \cite{DMV,BLMV}.
In \cite{DILM} this quantity was used to extract the angle $\alpha$ of the unitarity triangle from $B_d\to K^{0}\bar{K}^{0}$.
In the following section we review the formulae that allow to extract $T$ and $P$ from data and the theoretical input $\Delta$.

\section{Tree and Penguin Contributions}

\begin{figure}
\begin{center}
\psfrag{A}{\hspace{-1.5cm}$\Adir\lg(B_d\to K^{*0}\bar{K}^{*0})$}
\psfrag{BR}{\hspace{-1.5cm}$BR\lg(B_d\to K^{*0}\bar{K}^{*0})\times 10^6$}
\includegraphics[width=8cm]{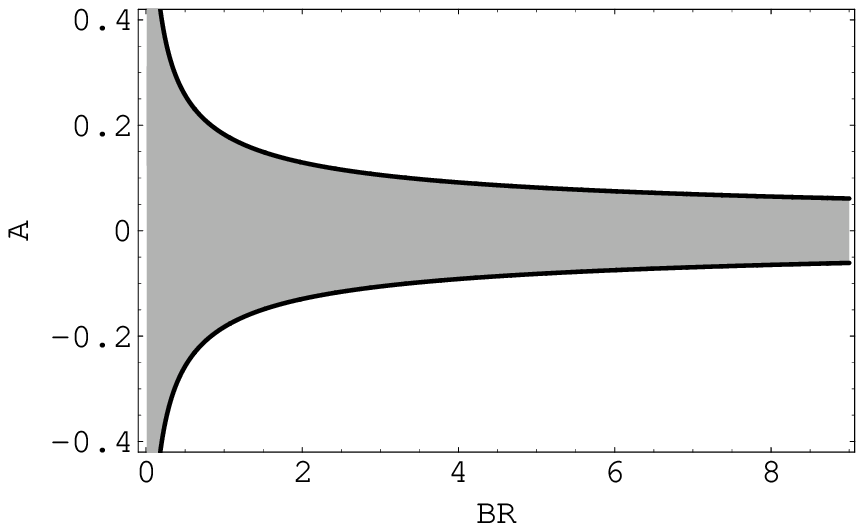}
\end{center}
\caption{Allowed values for the longitudinal direct CP asymmetry of
$B_d\to K^{*0}\bar{K}^{*0}$ in terms of its longitudinal branching ratio, according to the value of $\Delta_{K^*K^*}$.}
\label{fig1}
\end{figure}

We begin writing the two self-conjugated amplitudes in terms of tree and penguin contributions,
\eq{A=\lambda_u^{(D)*} T +\lambda_c^{(D)*} P\ ,\quad \bar{A}=\lambda_u^{(D)} T +\lambda_c^{(D)} P}
Now we put $T=P-\Delta$ and we square the amplitudes,
\eq{
\begin{array}{rcl}
|A|^2 & = & |\lambda_c^{(D)*}+\lambda_u^{(D)*}|^2\left| P + \frac{\lambda_u^{(D)*}}{\lambda_c^{(D)*}+\lambda_u^{(D)*}} \Delta \right|^2\\
&&\\
|\bar{A}|^2 & = & |\lambda_c^{(D)}+\lambda_u^{(D)}|^2\left| P + \frac{\lambda_u^{(D)}}{\lambda_c^{(D)}+\lambda_u^{(D)}} \Delta \right|^2
\end{array}}
But the squared amplitudes are directly related to observables,
\eqa{|A|^2&=&BR(1+\Adir)/g_{PS}\nonumber\\
|\bar{A}|^2&=&BR(1-\Adir)/g_{PS}}
where $g_{PS}$ is the usual phase-space factor. Neglecting the masses of the light mesons with respect to the $B$ mesons,
\eqa{
g_{PS}(B_d)&=&8.8\times 10^9\,{\rm GeV}^{-2}\nonumber\\
g_{PS}(B_s)&=&8.2\times 10^9\,{\rm GeV}^{-2}
}
for two non-identical particles in the final state. The resulting expressions are
\eqa{
\frac{BR(1+\Adir)/g_{PS}}{|\lambda_c^{(D)*}+\lambda_u^{(D)*}|^2}=
\left| P + \frac{\lambda_u^{(D)*}}{\lambda_c^{(D)*}+\lambda_u^{(D)*}} \Delta \right|^2\nonumber\\
\frac{BR(1-\Adir)/g_{PS}}{|\lambda_c^{(D)}+\lambda_u^{(D)}|^2}=\left| P + \frac{\lambda_u^{(D)}}{\lambda_c^{(D)}+\lambda_u^{(D)}} \Delta \right|^2
}
These are the equations for two circles in the complex $P$ plane, whose solutions are the two points of intersection. This will result
in a two-fold ambiguity in the determination of $P$ (and $T$). Before writing down the analytical solutions, notice that in order for
solutions to exist, the separation between the centers of these circles must be smaller than the sum of the radii but bigger than the
difference. This translates into a consistency condition between $BR$, $\Adir$ and $\Delta$:
\eq{|\Adir|  \le  \sqrt{\frac{\mathcal{R}_D^2\Delta^2}
{2\widetilde{BR}}\Big(
2-\frac{\mathcal{R}_D^2\Delta^2}{2\widetilde{BR}} \Big)}\label{cons}}
where $\widetilde{BR}\equiv BR/g_{PS}$ and $\mathcal{R}_D$ is a specific combination of CKM elements (see Table \ref{coeffs}).
This condition turns out to be highly nontrivial. For example, Fig.\ref{fig1} shows the allowed values for the longitudinal direct CP asymmetry of
$B_d\to K^{*0}\bar{K}^{*0}$ in terms of its longitudinal branching ratio. It can be seen that for $BR\gtrsim 3\times 10^{-6}$
the direct CP asymmetry is quite constrained.

The hadronic quantities $P$ and $T$ are then given by
\eqa{
Im[P] & = & \frac{\widetilde{BR}\,\Adir}{2 c_0^{(D)}\Delta}\nonumber\\
&&\nonumber\\
Re[P] & = & -c_1^{(D)}\,\Delta \pm
\sqrt{-Im[P]^2-\left(\frac{c_0^{(D)}\Delta}{c_2^{(D)}}\right)^2+\frac{\widetilde{BR}}{c_2^{(D)}}}\nonumber\\
&&\nonumber\\
T&=&P+\Delta
\label{eqTP}}
where the coefficients $c_i^{(D)}$ are again some specific combinations of CKM elements (see Table \ref{coeffs}).

\begin{table}
\begin{tabular}{cccc}
\hline\hline
$c_0^{(d)}$ & $c_1^{(d)}$ & $c_2^{(d)}$ &  $\mathcal{R}_d$\\
\hline
$\ -3.15\cdot 10^{-5}\ $  &  $\ -0.034\ $  &  $\ 6.93\cdot 10^{-5}\ $  &  $7.58\cdot 10^{-3}$  \\
\hline\hline
$c_0^{(s)}$    &    $c_1^{(s)}$   &   $c_2^{(s)}$  &  $\mathcal{R}_s$\\
\hline
$\ 3.11\cdot 10^{-5}\ $  &  $\ 0.011\ $ &  $\ 1.63\cdot 10^{-3}\ $ &  $1.54\cdot 10^{-3}$ \\
\hline\hline
\end{tabular}
\caption{Numerical values for the coefficients $c_i^{(D)}$ and $\mathcal{R}_D$ for $\gamma=62^\circ$.}
\label{coeffs}
\end{table}

Equations (\ref{eqTP}) allow to extract the hadronic parameters $T$ and $P$ from
experimental data on $BR$ and $\Adir$, information on sides of the unitarity triangle
and the weak phase $\gamma$, and the theoretical value for $\Delta$. This method is also
powerful because if no experimental information is available for $\Adir$, one can just
vary $\Adir$ over its allowed range in eq.(\ref{cons}). So in fact $T$ and $P$ can be extracted
from $BR$, $\Delta$ and CKM elements.

\section{$\sin{2\beta}$ from $B\to \phi K_s$}

Following the discussion in the previous section, the bounds for ${\rm Re}(T_{\phi K_s}/P_{\phi K_s})$ are given by
\eqa{
{\rm Re}\left( \frac{T_{\phi K_s}}{P_{\phi K_s}} \right)&\le&\,1+\left(-c_1^{(s)}+C(BR_{\phi K_s},\Delta_{\phi K_s}^d)\right)^{-1}\nonumber\\
{\rm Re}\left( \frac{T_{\phi K_s}}{P_{\phi K_s}} \right)&\ge&\,1+\left(-c_1^{(s)}-C(BR_{\phi K_s},\Delta_{\phi K_s}^d)\right)^{-1}\nonumber\\
C(BR,\Delta)&\equiv&\sqrt{-(c_0^{(s)}/c_2^{(s)})^2+(1/c_2^{(s)})\,\widetilde{BR}/\Delta^2}\qquad
\label{bounds}}
Introducing the numbers for the coefficients from Table \ref{coeffs}, the value or $\Delta_{\phi K_s}^d$ in Table \ref{table1}
and the latest experimental value for the branching ratio \cite{HFAG}
\eq{BR(B_d\to \phi K_s)_{\rm exp}=8.3^{+1.2}_{-1.0}\times 10^{-6}}
we get the following bounds for $\Delta S_{\phi K_s}$,
\eq{0.03<\Delta S_{\phi K_s}<0.06}

\section{$\sin{2\beta_s}$ from $B_s\to VV$}

Equations (\ref{bounds}) apply as well to $B_s\to VV$. Here we focus on longitudinal polarizations
for which the numerical values of $\Delta$ are under control. As mentioned above, penguin
mediated $B_s\to VV$ decays measure $\sin{2\beta_s}$, but no experimental information is
available yet for CP asymmetries in $B_s$ decays. There is, though, an experimental number for the
$B_s\to \phi\phi$ branching ratio \cite{HFAG},
\eq{BR(B_s\to\phi\phi)_{\rm exp}=14^{+8}_{-7}\times 10^{-6}}
If we suppose that the longitudinal polarization fraction is $f_L^{\phi\phi}\sim 50\%$
as QCDF suggests \cite{beneke}, then we find
\eq{0.006\le \Delta S_{\phi\phi}\le 0.072}

The decay $B_s\to K^{*0}\bar{K}^{*0}$ is more appropriate because $\Delta_{K^*K^*}$ is under a much better numerical control,
but there is no experimental value for the branching ratio. For the sake of illustration, we just mention that for
$BR\lg(B_s\to K^{*0}\bar{K}^{*0})\sim (30-40)\times 10^{-6}$, one gets
\eq{0.037\le \Delta S_{K^*K^*}\le 0.051}

\section{Other angles from Data and $\Delta$}

There are also some expressions that can be written that relate directly branching ratios and asymmetries
to other angles of the unitarity triangle through the quantity $\Delta$ \cite{BVV}. These expressions do not require
any CKM angle as an input, just sides of the unitarity triangle. The experimental input is minimized by measuring
$\mathcal{A}_{\Delta\Gamma}$, which can be extracted from the time-dependent untagged rate \cite{Dighe,fleischerMatias}.
Then, in the case of a $B_d$ meson decaying through a $b\to D$ process ($D=d,s$),
\eq{
\sin^2{\alpha}=\frac{\widetilde{BR}(1-\mathcal{A}_{\Delta\Gamma})}{2|\lambda_u^{(D)}|^2|\Delta|^2}\ ;\
\sin^2{\beta}=\frac{\widetilde{BR}(1-\mathcal{A}_{\Delta\Gamma})}{2|\lambda_c^{(D)}|^2|\Delta|^2}
}
and in the case of a $B_s$ meson decaying through a $b\to D$ process ($D=d,s$),
\eq{
\sin^2{\beta_s}=\frac{\widetilde{BR}(1-\mathcal{A}_{\Delta\Gamma})}{2|\lambda_c^{(D)}|^2|\Delta|^2}}

\section{Predictions for $B_s\to K^*K^*$}

Assuming no new physics contributing to $B_d\to K^{*0}\bar{K}^{*0}$, we can also give SM predictions for the branching ratios
and CP asymmetries of its U-spin partner $B_s\to K^{*0}\bar{K}^{*0}$ \cite{BVV}. The inputs are $\Delta_{K^*K^*}^d$, SU(3) breaking
from QCDF, and the experimental value for $BR(B_d\to K^{*0}\bar{K}^{*0})$. While there is still no experimental information
on this branching ratio, it is remarkable that for $BR(B_d\to K^{*0}\bar{K}^{*0})\gtrsim 5 \times 10^{-7}$ the results are almost
insensitive to it. Taking $\gamma=(62\pm 6)^\circ$ and $2\beta_s=-2^\circ$, we find
\eqa{
&&\left(\frac{BR\lg(B_s\to K^{*0}\bar{K}^{*0})}{BR\lg(B_d\to K^{*0}\bar{K}^{*0})}\right)_{\sss SM}=17\pm 6\label{br}\\
&&\nonumber\\
&&\Adir\lg(B_s\to K^{*0}\bar{K}^{*0})_{\sss SM}=0.000\pm 0.014\label{as}\\
&&\nonumber\\
&&\Amix\lg(B_s\to K^{*0}\bar{K}^{*0})_{\sss SM}=0.004\pm 0.018.
}

\section{Conclusions}

To conclude, we would like to comment on the relevance of the proposal. First, the applicability of the approach has to be checked
individually for each mode: it can only be applied to those decays for which $\Delta$ receives no contributions from annihilation or
hard spectator scattering graphs, such as $B\to K^{(*)}K^{(*)}$, $B_d\to\phi K^{(*)}$, $B^+\to \pi^+\phi$, etc.
The predictions derived in this way include most of the long distance physics, which is contained inside the experimental
input. The used theoretical input is minimal, and is the most reliable input that QCDF can offer, free from the troublesome IR-divergencies.
Moreover, the theoretical error is under control and is likely to be reduced in the near future due to, for example,
the fast progress that is taking place in lattice simulations.

There is, however, an honest criticism due to the fact that some long distance effects that are controversial in QCDF, are also
absent here. The prominent one is the contribution from the charming penguins, which could easily account for a significant discrepancy
between theory and experiment that would not be due to NP. A recent example of this is given in \cite{Stewart}.

\section*{Acknowledgments}

I would like to thank Joaquim Matias for a careful reading of the manuscript, and to Sebastian Descotes-Genon for
a nice and dynamical collaboration. I am also indebted to Gudrun Hiller for helpful criticism and discussions on the
part concerning $\sin 2\beta$.
This work was supported in part by the EU Contract No. MRTN-CT-2006-035482, ``FLAVIAnet''.

\end{document}